\documentclass{aa}  
\usepackage{graphicx}
\usepackage[varg]{txfonts}
\usepackage{color}
\usepackage{natbib,twoopt}
\usepackage[colorlinks=true,
            linkcolor=blue,
            urlcolor=blue,
            citecolor=blue]{hyperref}
\bibpunct{(}{)}{;}{a}{}{,} 

\begin{document} 
	\title{Modelling polarized light from dust shells surrounding asymptotic giant branch stars}
	\subtitle{}

   	\author{E. Aronson
   		\inst{1}
   		\and S. Bladh
   		\inst{2}
   		\and S. Höfner
   		\inst{1}}
   	
   	\institute{Division of Astronomy and Space Physics, Department of Physics and Astronomy, Uppsala University, Box 516, SE-751 20 Uppsala, Sweden.\\
   		\email{erik.aronson@physics.uu.se}
   		\and
   		Dipartimento di Fisica e Astronomia Galileo Galilei, Università di Padova, Vicolo dell’Osservatorio 3, 35122 Padova, Italy
   		\\
   	}
   \date{Received January 25, 2017; accepted May 18, 2017}
%
 
 	\abstract
 	{Winds of asymptotic giant branch (AGB) stars are commonly assumed to be driven by radiative acceleration of dust grains. For M-type AGB stars, the nature of the wind-driving dust species has been a matter of intense debate. A proposed source of the radiation pressure triggering the outflows is photon scattering on Fe-free silicate grains. This wind-driving mechanism requires grain radii of about 0.1 -- 1 micron in order to make the dust particles efficient at scattering radiation around the stellar flux maximum. Grain size is therefore an important parameter for understanding the physics behind the winds of M-type AGB stars.} 
 	{We seek to investigate the diagnostic potential of scattered polarized light for determining dust grain sizes.}
 	{We have developed a new tool for computing synthetic images of scattered light in dust and gas shells around AGB stars, which can be applied to detailed models of dynamical atmospheres and dust-driven winds.}
 	{We present maps of polarized light using dynamical models computed with the DARWIN code. The synthetic images clearly show that the intensity of the polarized light, the position of the inner edge of the dust shell, and the size of the dust grains near the inner edge are all changing with the luminosity phase. Non-spherical structures in the dust shells can also have an impact on the polarized light. We simulate this effect by combining different pulsation phases into a single 3D structure before computing synthetic images. An asymmetry of the circumstellar envelope can create a net polarization, which can be used as diagnostics for the grain size. The ratio between the size of the scattering particles and the observed wavelength determines at what wavelengths net polarization switches direction. If observed, this can be used to constrain average particle sizes.}
 	{}
 	
 	\maketitle

\section{Introduction}
Winds of asymptotic giant branch (AGB) stars are commonly assumed to be driven by radiative acceleration of dust grains. The pulsation of these stars give rise to shock waves, which propagate through the atmosphere and intermittently lift the upper layers to a few stellar radii. As the compressed gas cools down while moving away from the stellar photosphere, dust formation can start. The dust particles experience radiation pressure from stellar photons, thereby accelerating these grains outwards. Momentum from these accelerated grains is transferred to the surrounding gas through collisions, and thus an outflow of gas and dust from the stellar photosphere is triggered. 

For AGB stars with a carbon-rich atmospheric chemistry ($\mathrm{C/O > 1}$) there is a broad consensus that the observed outflows are driven by amorphous carbon grains. For M-type AGB stars ($\mathrm{C/O < 1}$), however, the nature of the wind-driving dust species has been a matter of intense debate. Magnesium-iron silicates (olivine- and pyroxene-type materials) seem to be good candidates, considering the prominence of silicate features in mid-IR spectra and the comparatively high abundances of the constituent elements (Si, Mg, Fe, and O). Detailed models, however, show that silicate grains in the close vicinity of an AGB star, in the layers where the wind originates, can only have a very low content of Fe to avoid destruction by radiative heating \citep{Woitke2006AA}. This leads to low levels of absorption at visual and near-IR wavelengths, corresponding to a radiative pressure that is insufficient to drive an outflow. 

Using detailed atmosphere and wind models, \citet{Hofner2008AA}, \citet{Bladh2013AA, Bladh2015AA}, and \citet{Hoefner2016AA} have therefore investigated an alternative source of radiation pressure, i.e. scattering of stellar photons on Fe-free silicate grains. This wind-driving mechanism requires grain radii of about 0.1 -- 1 micron to make the dust particles efficient at scattering radiation in the near-IR wavelength region where the stellar flux peaks.

Determining dust particle sizes in the atmospheres and wind formation regions of AGB stars is not a straightforward task. At visual and near-IR wavelengths, where there is a diagnostic potential for the expected grain sizes, the central star is significantly brighter than the dust shell, which -- in the best-case scenarios -- is separated by only a few tens of milliarcseconds from the stellar surface. However, the brightness of dust shells in the visual and near-IR region is dominated by scattered stellar light, which has a high degree of linear polarization. Therefore, by observing in linearly polarized light the photons scattered by the dust shell can be distinguished from the direct stellar radiation. 

Observations by \citet{Norris2012Natur} provided strong indications that dust particles with sizes where photon scattering can drive an outflow actually exist around AGB stars. These observations used aperture masking polarimetric interferometry \citep{Norris2012SPIE} on three nearby AGB stars (W Hya, R Dor, and R Leo). A linearly polarized signal was detected approximately at a distance of 2 stellar radii from the centre of the star, which was interpreted as a geometrically thin layer of dust particles scattering light towards the observer. Fitting the observed wavelength-dependent fraction of scattered light to the scattering efficiency of dust grains of various sizes gave a best fit for grains with radii around 0.3 $\mathrm{\mu m}$. 

More recently, resolved direct images of polarized light in the close vicinity of the AGB stars W Hya and R Dor were obtained with the Very Large Telescope (VLT), using the Spectro-Polarimetric High-contrast Exoplanet REsearch (SPHERE) - Zurich imaging polarimeter (ZIMPOL). \citet{ohnaetal16a} have detected three clumpy dust clouds at about 2--3 stellar radii around W Hya, which are comparable to structures resulting from large-scale convection in the 3D star-in-a-box models of \citet{FreyHoef2008}. The analysis of the maps of the observed polarized intensity has indicated that they represent an optically thin medium with density enhancements of about a factor of 4 in the clouds and grain radii of 0.4--0.5 microns. A second-epoch measurement, which was taken about eight months later at minimum light \citep{ohnaetal16b}, revealed changes in the morphology of the clumps, and in grain size (showing a dominant grain population with radii of about 0.1 microns), compared to the first observation that was taken at pre-maximum light. These changes are consistent with an interpretation of grains forming around the luminosity minimum and growing during the phases of increasing luminosity; this also fits well with detailed dynamical atmosphere and wind models. \citet{khouetal16} have obtained high-angular-resolution images of the AGB star R Dor at two epochs, demonstrating that the size and morphology of surface patterns change significantly in less than two months; this is compatible with convection-induced structures found in 3D models. These authors have interpreted these changes as variability in the excitation and/or density of TiO. They also observe light scattered by dust grains close to the star and a strong decrease in the dust density further out, which  can be interpreted as rapidly falling densities due to wind acceleration or a change in mass loss rate. 

In this paper we explore interpretations of polarized signatures from AGB stars from a theoretical point of view. In contrast to the simple, empirical structures used in the observational papers discussed above, our results are based on structures of the extended atmosphere and dusty circumstellar envelope, which are obtained with detailed, time-dependent dynamical atmosphere and wind models. Polarimetric signatures from the gas and dust shells are modelled and observables quantities that provide diagnostics of dust grain sizes are discussed.


\section{Modelling: Methods and parameters}
\subsection{Wind models} \label{windModels}
The atmospheric structures, for which scattered light is modelled, are snapshots taken from the dynamical models of M-type AGB stars presented in \citet{Hofner2008AA} and \citet{Bladh2013AA}. In these detailed models of the atmosphere and wind of AGB stars the mass loss is driven by a pulsation-enhanced dust-driven wind, where atmospheric levitation by pulsation-induced shock waves is followed by radiative acceleration of newly formed dust grains.  

The dynamical models, produced with the DARWIN code \citep{Hoefner2016AA}, are spherically symmetric with an inner boundary situated just below the photosphere and an outer boundary at about 20-30 stellar radii for models that develop a wind. The parameters that define the hydrostatic initial models are stellar mass $M_{\star}$, stellar luminosity $L_{\star,}$ and the effective temperature of the star $T_{\star}$ along with elemental abundances. The variability of the star is simulated by a sinusoidal variation of velocity and luminosity at the inner boundary of the model, characterized by a pulsation period $P$ and the amplitudes of velocity and luminosity variations (parameters $u_{\mathrm{p}}$ and $f_{\rm L}$; see \citet{Hoefner2016AA} for details). The dynamical structure of the atmosphere, including the formation of shock waves and the development of stellar winds, is produced by simultaneously solving the equations of hydrodynamics, frequency-dependent radiative transfer, and grain growth. The abundances of atoms and molecules in the outer cool layers of the atmosphere are given by chemical equilibrium and the models include frequency-dependent opacities of both gas and dust in the radiative transfer to achieve realistic density and temperature structures. Dust formation is modelled with a time-dependent description for growth and evaporation of Fe-free silicate ($\mathrm{Mg_{2}SiO_{4}}$) grains. As there is currently no well-developed nucleation theory for oxygen-rich atmospheres, the efficiency of grain formation is modelled by setting the number of seed particles per hydrogen atom ($n_{gr}/n_H$) as an input parameter. These seed particles start to grow when the thermodynamic conditions favour grain growth. As a consequence, the models produce a uniform grain size for all dust particles at a given distance from the stellar surface, at a given time.

As a direct result, the DARWIN models produce time series of snapshots of the radial structure of the atmosphere and wind. The models thereby provide information about properties such as velocity, temperature, density, and dust grain size as a function of radial distance and time. Selected snapshots are further processed in an a posteriori radiative transfer calculation to produce spectra and photometry, using the opacities from the COMA code \citep{BADiss00, Aringer09}. Detailed information about the partial pressures of individual atoms and molecules (assuming chemical equilibrium) as a function of radial distance is also produced in this calculation. Three different atmosphere models are used in this paper, where the seed particle abundances ($n_{gr}/n_H$) differs and all other input parameters are similar. This generates dust shells with differently sized dust grains; see Table \ref{modelpar}.

\begin{table}[h]
	\caption{Model parameters. \newline
		Input parameters: stellar luminosity $L_{\star}$, effective temperature $T_{\star}$, and seed particle abundance (relative to hydrogen) $n_{\mathrm{gr}}/n_{\mathrm{H}}$. Results: terminal wind velocity $\langle u \rangle$ and average dust grain radius $\langle a_{\mathrm{gr}} \rangle$.}
	\centering
	\begin{tabular}{lllllll}
		\hline
		Model & $M_{\star}$ & $L_{\star}$ & $T_{\star}$ & $n_{\mathrm{gr}}/n_{\mathrm{H}}$ & $\langle u \rangle$ & $\langle a_{\mathrm{gr}} \rangle$ \\
		& (M$_\odot$) & (L$_\odot$) & ($\mathrm{K}$) & & ($\mathrm{km/s}$) & ($\mathrm{\mu m}$)\\
		\hline
		B1 & 1.0 & 7000 & 2700 & $\mathrm 3\times10^{-16}$ & 4 & 0.66\\
		B2 & 1.0 & 7000 & 2700 & $\mathrm 1\times10^{-15}$ & 7 & 0.45\\
		B3 & 1.0 & 7000 & 2700 & $\mathrm 3\times10^{-15}$ & 11 & 0.34\\
		\hline
	\end{tabular}
	\label{modelpar}
\end{table}

\subsection{Polarimetric radiative transfer} \label{radTrans}
In this section we describe the modelling of scattered light in the dust-enriched extended atmosphere of an AGB star. We calculate the state of polarization of scattered light as a function of wavelength and position within the atmosphere, which we then use to synthesize observations in polarized light with high spatial resolution.

\subsubsection{Scattering particles}
In order to fully include polarization in the calculations, we use the four Stokes parameters to describe intensity,
\begin{equation} \label{stokes} 
I = \begin{bmatrix}
i\\q\\u\\v
\end{bmatrix}.
\end{equation}
All scattering processes are described by multiplication of $I$ with a 4 by 4 scattering matrix ($F$). This matrix has the property that multiplication with incoming intensity and single scattering albedo ($a$) gives the intensity (including state of polarization) of the scattered light. Scattering matrices are typically functions of the scattering angle $\theta$, defined as the angle between the direction of propagation of the light before and after the scattering event,
\begin{equation} \label{ScatterEq}
I_{sca} = a F(\theta)I_{in}.
\end{equation}

The first step in modelling scattered light is to create scattering matrices for particles in the atmosphere. We divide particles into two categories: small particles such as atoms and molecules and large particles such as dust grains. Small particles (for which we use the prefix $m$) are significantly smaller than the wavelength of the light. For these we use Rayleigh scattering theory to describe the scattering process. Large particles (prefix $d$) are comparable in size to the wavelength of the light and Mie scattering theory is used to describe scattered light in that case.

Rayleigh and Mie scattering matrices are calculated as given by \citet{Hansen1974SSRv}. For Mie Scattering we use the light scattering code BHMIE, based on \citet{Bohren1983asls}.To ensure scattering matrices are correctly computed, we compared properties of our scattering matrices with the same properties generated from the light scattering code MiePlot. Both codes produced identical results. A comparison is shown in fig.~\ref{S21_all}.

\subsubsection{Atmospheric shells}
The extended atmosphere is discretized into 300 homogeneous spherically symmetric shells of equal thickness. Within each shell the properties of gas and dust are known and assumed to be constant. We set the boundary of the innermost shell at 1.2 stellar radii ($R_{\star}$), as there are no dust particles closer to the star than this and the outer boundary at $10\,R_{\star}$, since outside of this region the density is too low to produce a significant amount of scattered light.

We combine the scattering matrices of individual species or sizes of particles into scattering matrices for entire shells, using the methods described by \citet{deHaan1987AA}. For a homogeneous atmosphere, taking only single scattering into account, this can be carried out using
\begin{align} \label{singlescatter}
	& F(\theta, Z, \lambda) = \\ \nonumber
	& \quad { \sum_{k} \big[ b_{sca}^{m,k}(Z, \lambda) \cdot F^{m,k}(\theta, \delta(k)) \big]
		+ \sum_{l} \big[b_{sca}^{d,l}(Z, \lambda) \cdot F^{d,l}(\theta , x , y)  \big] \over
		\sum_{k} b_{sca}^{m,k}(Z, \lambda) + \sum_{k} b_{sca}^{d,l}(Z, \lambda)}
\end{align}
as described by \citet{Stam2000JQSRT}. Here $Z$ denotes the index of the atmospheric shell (corresponding to distance from star), $b$ the single scattering optical thickness, and $F$ the scattering matrices ($F^m$ for Rayleigh and $F^d$ for Mie scattering). Summation is carried out over the different molecular species ($k$) and sizes of dust grains ($l$) found in the given shell.

The single scattering optical thickness for molecules ($b_{sca}^m$) is given by
\begin{equation} \label{OTmole}
b_{sca}^m(Z, \lambda)=\eta(Z) { 24 \pi ^3 N_{av} (n^2 (\lambda)-1)^2 (6 + 3 \delta) \over 
	\lambda ^4 N^2_L R(n^2 (\lambda)+2)^2 (6 - 7 \delta ) }
\int\limits_{Z_i}^{Z_j} {p(z) \over T(z)} dz ,\end{equation}
where $N_{av}$ is the Avagado constant, $N_L$ is the Loschmidt constant, $R$ is the gas constant, $n$ is the refraction index, $\delta$ is the depolarization factor, $p(z)$ is the pressure as a function of distance from the star,  and $T(z)$ is temperature as a function of distance from the star. To incorporate the number density of different atoms and molecules, the expression is multiplied with the volume mixing ratio $\eta(Z)$ of the atomic or molecular species (assumed to be constant within the shell). The integration over height is carried out assuming exponential decrease in $P/T$ through the shell, where $P(z_i)$, $T(z_i)$, $P(z_j)$, and $T(z_j)$ correspond to tabulated values at the top and bottom of the shell.

The single scattering optical thickness for dust grains ($b_{sca}^d$) are given by
\begin{equation} \label{OTaero}
b_{sca}^d (Z, \lambda)= N(r,Z)\cdot(z_i - z_j)\cdot \sigma _{sca} (r, \lambda).
\end{equation}
Here $N(r,Z)$ is the number density of particles with radius $r$ in shell $Z$, which is assumed to be constant within the shell; $\sigma _{sca}$ is the scattering cross section of particles with radius $r$, calculated using the BHMIE code.

With both scattering and absorption optical thickness of molecules and dust grains, see Eq.~\ref{AbsMole} and Eq.~\ref{AbsAero}, we can compute the combined optical thickness of an atmospheric shell (Eq.~\ref{OTlayer}),
\begin{equation} \label{AbsMole}
b^m_{abs}(Z, \lambda) = \eta (Z) {N_{av} \sigma _{abs}^m (Z, \lambda) \over R}
\int\limits_{Z_i}^{Z_j} {p(z) \over T(z)} dz
.\end{equation}
The integral is evaluated assuming exponential decrease through the shell. The molecular absorption cross section $\sigma _{abs} $ is assumed to be constant within the shell.
\begin{equation} \label{AbsAero}
b^d_{abs}(Z, \lambda)= N(r,Z)\cdot(z_i - z_j)\cdot \sigma _{abs} (r, \lambda)
\end{equation}

\begin{align} \label{OTlayer}
	& b(Z, \lambda) = \\ \nonumber
	& \sum_{k} b_{sca}^{m,k}(Z, \lambda) + 
	\sum_{k} b_{abs}^{m,k}(Z, \lambda) + 
	\sum_{l} b_{sca}^{d,l}(Z, \lambda) + 
	\sum_{l} b_{abs}^{d,l}(Z, \lambda).
\end{align}

\subsubsection{Radiative transfer}
\begin{figure} 
	\centering
	\includegraphics[width=\hsize]{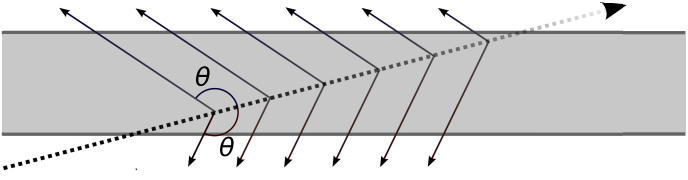}
	\caption{Possible paths for light moving through an atmospheric shell. The transmission matrix describes light emerging at the upper boundary of this shell and the reflection matrix describes light emerging from the lower boundary of this shell.}
	\label{transMatrix}
\end{figure}

With scattering matrices for all atmospheric shells, we can now perform the radiative transfer. In the first step we consider each shell individually. For each shell, we create reflection and transmission matrices \citep{deHaan1987AA}. These matrices have the property that when multiplied with the incoming light arriving at the bottom of the atmosphere, the product is the light emerging from the upper (reflection matrix) or lower (transmission matrix) boundary of the layer. Thus, given the incident light on a shell and its angle to the local zenith, we can calculate the intensity (including state of polarization) of the outgoing light at a given angle to local zenith. The separation into reflection and transmission matrix is needed owing to the different paths light has to take to reach either upper or lower boundary of the shell; see Fig.~\ref{transMatrix}.

\begin{figure}
	\centering
	\includegraphics[width=\hsize]{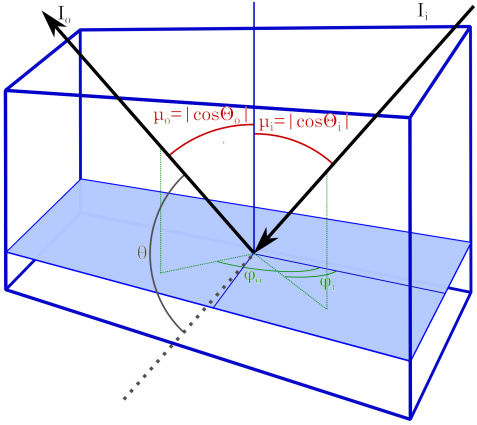}
	\caption{Direction of propagation of incoming and outgoing light is defined by $\mu$ and $\varphi$.}
	\label{anglesInAtmosphere}
\end{figure}

For transmission and reflection matrices, the direction of propagation of the incoming and outgoing light is needed. This is not uniquely defined by scattering angle, and we therefore introduce $\mu$ and $\varphi$ to define direction of propagation; $\mu = \vert \cos \Theta \vert$ is the angle between the direction of propagation of the light and the local zenith, where $\mu_o$ is for the outgoing light and $\mu_i$ is for incoming light. The azimuth angle $\varphi_o$ (outgoing) and $\varphi_i$ (incoming) are measured from an arbitrary plane containing the local zenith and clockwise when looking downwards; see Fig.~\ref{anglesInAtmosphere}. Because of rotational symmetry only the difference in azimuth angles ($\varphi = \varphi_i-\varphi_o$) is relevant.

When multiple scattering is neglected and no internal light sources are present inside the shell, the intensity of light at optical depth $\tau$ in an atmospheric shell can be calculated using 
\begin{align} \label{Idown}
	& I_{down} (\tau,\mu_o,\varphi, \lambda) = 
	\Bigg( {1 \over \pi}
	\int\limits_{0}^{1} \mu_i d\mu_i 
	\int\limits_{0}^{2 \pi} d\varphi \cdot
	\bigg[ {1 \over 4 \mu_o \mu_i}
	\int\limits_{0}^{\tau} d\tau ' \cdot \nonumber \\       
	& \indent e^{-\tau ' \over \mu_i}
	e^{-(\tau -\tau ')\over \mu_o}
	a(\tau ')
	\times F(\theta(\mu_i, \mu_o, \varphi), \lambda)
	\bigg]  + e^{-\tau \over \mu_o}  \Bigg) 
	\cdot I_i(0 ,\mu_i, \varphi, \lambda)
\end{align}
\begin{align}\label{Iup}
	& I_{up}(\tau ,-\mu_o, \varphi, \lambda) = 
	\Bigg( {1 \over \pi} 
	\int\limits_{0}^{1} \mu_i d\mu _i 
	\int\limits_{0}^{2 \pi} d\varphi        \cdot
	\bigg[ {1 \over 4 \mu_o \mu_i}                          
	\int\limits_{\tau}^{b} d\tau '  \cdot \nonumber \\      
	& \indent e^{-\tau ' \over \mu_i}
	e^{-(\tau ' -\tau)\over \mu_o}
	a(\tau ') 
	\times F(\theta(\mu_i, -\mu_o, \varphi), \lambda)
	\bigg] \Bigg) 
	\cdot I_i(\tau,\mu_i, \varphi, \lambda).
\end{align}
Here $I_{down}$ is the intensity of light moving downwards at optical depth $\tau$ and at an angle given by $\mu_o$ and $\varphi$, and likewise $I_{up}$ is the intensity of light moving upwards at a direction given by $\mu_o$ and $\varphi$. The integration over $\tau '$ is to account for reflections along the depth of the shell adding to the intensity, which for light moving downwards is from $0$ (top of the layer) to $\tau$, and for light moving upwards from $b$ (bottom of layer) to $\tau$. The exponentials give the attenuation when the light moves through the atmosphere at the angle given by $\mu_i$ (before scattering) or $\mu_o$ (after scattering).

The integration within the square brackets can be evaluated analytically for a homogeneous atmospheric shell; $a$ is considered to be constant within the shell. We call these matrices reflection matrix $R$ when light entering the boundary of the layer is scattering back towards the same boundary ($0^\circ \leq \Theta_o < 90^\circ$) and transmission matrix $T$ when light is emerging at the opposite boundary ($90^\circ \leq \Theta_o \leq 180^\circ$). Integration over $\tau$ is carried out from $0$ to $b$ in both cases, i.e. we calculate the intensity of light at the top or bottom of the shell for light moving upwards and downwards, respectively, as follows:\begin{equation} \label{Rmatrix}
R(\mu_i,\mu_o ,\varphi, \lambda) = 
{a \over 4(\mu_o + \mu_i)} \bigg(
1 - e^{{-b \over \mu_o}+{-b \over \mu_i}} \bigg)
\times F(\theta(\mu_i,\mu_o ,\varphi), \lambda);
\end{equation}
if $ \mu_o \neq \mu_i$,
\begin{equation} \label{Tmatrix}
T(\mu_i,\mu_o ,\varphi, \lambda) = 
{a \over 4(\mu_o - \mu_i)} \bigg(
e^{-b \over \mu_o} - e^{-b \over \mu_i}\bigg)
\times F(\theta(\mu _0,\mu ,\varphi), \lambda);
\end{equation}
if $ \mu_o = \mu_i$,
\begin{equation} \label{T2matrix}
T(\mu_i,\mu_o ,\varphi, \lambda) = 
{a \over 4\mu_o ^2} 
e^{-b \over \mu_o}
\times F(\theta(\mu_i,\mu_o ,\varphi), \lambda).
\end{equation}

\begin{figure} 
	\centering
	\includegraphics[width=\hsize]{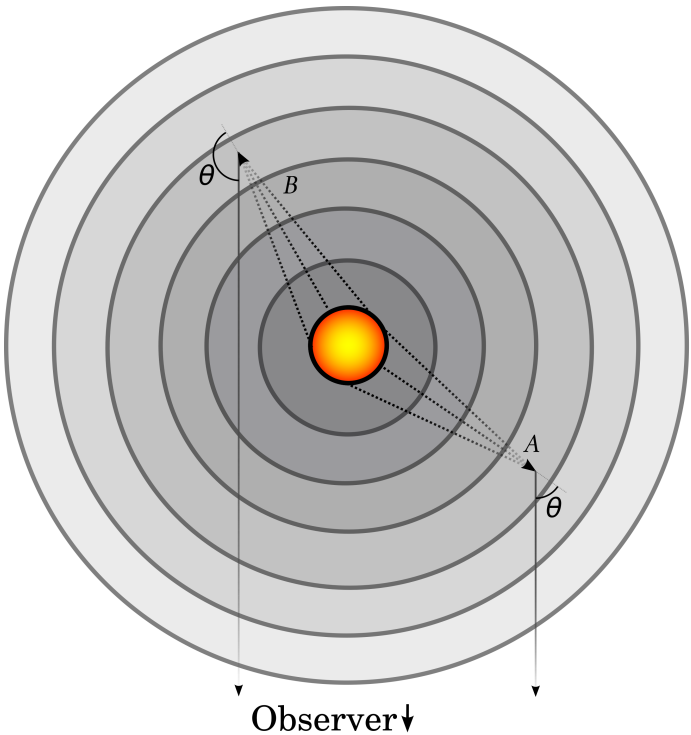}
	\caption{Sketch of the scheme for radiative transfer calculations. Stellar light moves outwards from the star to a point at which some of the light is scattered towards a distant observer. From this point light travels through the outer (case A) or first through the inner and then the outer (case B) before it reaches the boundary of our model.}
	\label{radiativetransfer}
\end{figure}

With transmission and reflection matrices for all shells and all combinations of angles for incoming and outgoing light, we can now perform radiative transfer calculations for the entire extended atmosphere. We consider only a plane of the atmosphere (the 3D structure is reintroduced in Sec.~\ref{syntObsCreation}), and calculate the amount of scattered light from each point on this plane for all shells. When calculating the amount of scattered light reaching the observer from some point in the atmosphere, there are two distinct cases, A and B in Fig.~\ref{radiativetransfer}. For forwards scattered light (case A), we first trace stellar light through the inner shells (using their transmission matrices where incoming and outgoing angles are the same) to the scattering point. We then compute the intensity of light scattered in a direction towards the observer using the transmission matrix of this shell and the relevant incoming and outgoing angles. Light is then traced through the outer layers as it travels towards the outer boundary of our model using the transmission matrices of the shell, where incoming and outgoing angles are the same. For backscattered light (case B), we trace stellar light through the inner shells to the scattering point and compute the intensity of light scattered in a direction towards the observer using the reflection matrix of this shell. Light is then traced through first the inner and then the outer shells as it travels towards the observer. Light that is absorbed or scattered in directions not towards the observer is assumed to be lost, i.e. no multiple scattering is included, which implies that the results are only valid for optically thin atmospheres. Using this approach should be acceptable when modelling typical M-type ABG stars. Their spectra shows very little absorption from dust in the visual and near-infrared (near-IR) wavelength regions and very little circumstellar reddening, resulting in an optically thin atmosphere.

\subsubsection{Synthesizing observations} \label{syntObsCreation}
These calculations generate the scattered light, including state of polarization, from each point within a plane of the shell. From this data a 1D image is constructed by combining scattered light along the same line of sight, but emerging from different shells. Since the extended atmosphere is assumed to be spherically symmetric, we can use this to construct a 2D image by postulating that the intensity is a function of distance from the centre of the star. We remove the intensity emerging from closer than $1.2\,R_{\star}$ (projected distance as seen by a distant observer); i.e. we remove the direct stellar light. 

\begin{figure}
	\centering
	\includegraphics[width=\hsize]{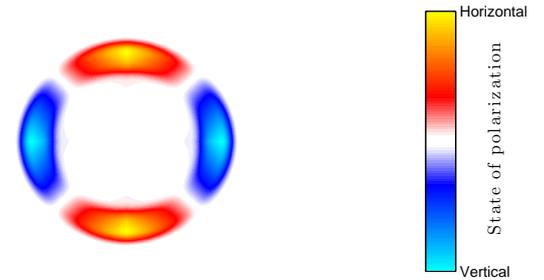}
	\caption{Synthesized image in linearly polarized light ($\lambda = 2.3\,\mathrm{\mu m}$) of a simplified dust shell model (a single thin layer containing dust particles with a grain radius of $0.3\,\mathrm{\mu m}$ and no gas). Blue represents the horizontally polarized light. Red represents the vertically polarized light. Just as expected, we observe bright arcs of polarized light where dust grains are situated.}
	\label{simplemodel}
\end{figure}

This radiative transfer scheme was tested on a simple extended atmosphere with no gas and only a single thin layer of dust grains situated at $2.5\,R_{\star}$; these dust grains all have a grain radius of $0.3\,\mu$m, which is similar to the set-up used by \citet{Norris2012Natur}. The results from this test is shown in Fig.~\ref{simplemodel}. As expected, when the circumstellar dust shell is viewed in linearly polarized light one can observe bright arcs at the distance where dust particles are present. The intensity outside these arcs is zero and the intensity inside the arcs is decreasing towards the star; the intensity from the inner parts comes from backwards and forwards scattered light.

\section{Results}
\subsection{Synthetic polarized images of detailed models}
\begin{figure}
	\centering
	\includegraphics[width=\hsize]{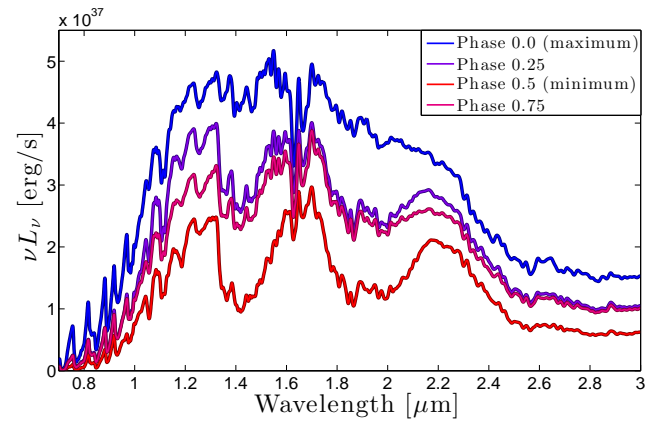}
	\caption{Spectral energy distribution for the model B2 at four different pulsation phases.}
	\label{spectra}
\end{figure}

\begin{figure} 
	\centering
	\includegraphics[width=\hsize]{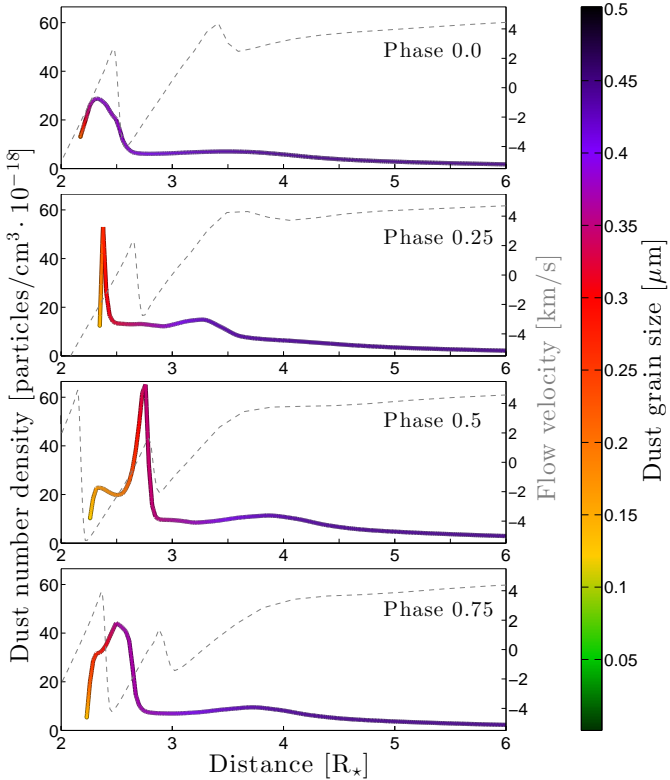}
	\caption{Density structure of the dust component as a function of radial distance for the model B2, taken at four different pulsation phases (same as in Fig.~\ref{spectra}). Colour-coded solid lines show density and dust grain size. Dashed grey lines show the flow velocity.}
	\label{densityPlot}
\end{figure}

\begin{figure}
	\centering
	\includegraphics[width=\hsize]{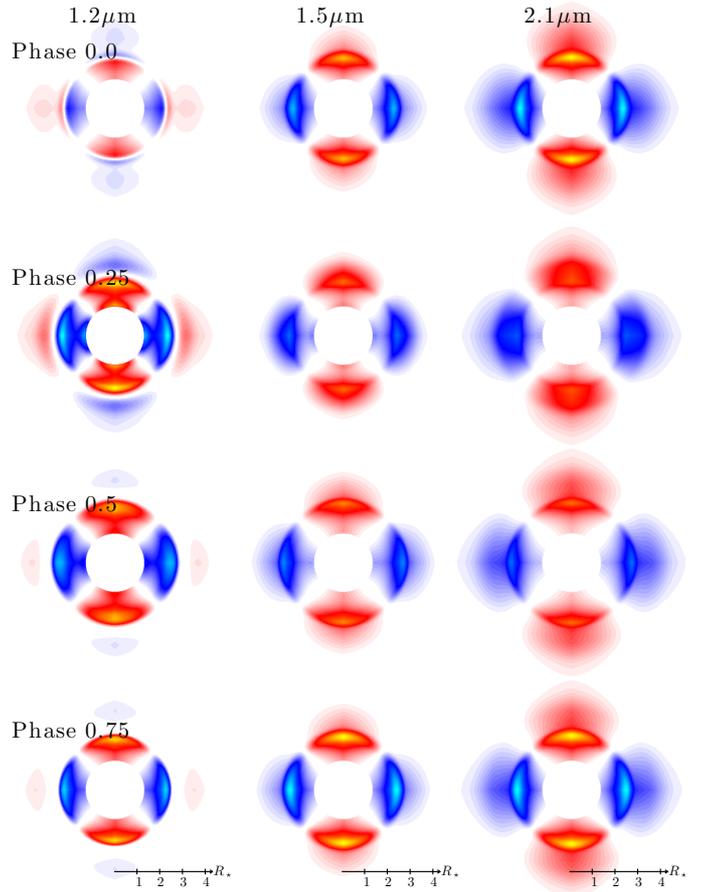}
	\caption{Linearly polarized light from the spherically symmetric model B2. Colour coding the same as in Fig.~\ref{simplemodel}. Scattered light computed at the maximum phase (0.0), minimum phase (0.5), and intermediate phases (0.25 and 0.75) is shown.}
	\label{B2allphase}
\end{figure}

The radiative transfer method described in Sect.~\ref{radTrans} is applied to snapshots of detailed wind models taken at different pulsation phases. We use model B2 for in-depth studies (see Tab.~\ref{modelpar}). Synthetic spectra for this model at four different phases are shown in Fig.~\ref{spectra}. The corresponding density and grain-size profiles, as a function of radial distance from the star, as well as the flow velocities, are shown in Fig.~\ref{densityPlot}. Images in linearly polarized light were synthesized for the same four phases and in three different wavelength regions ($1.2\,\mathrm{\mu m}$, $1.5\,\mathrm{\mu m,}$ and $2.1\,\mathrm{\mu m}$). Results are shown in Fig.~\ref{B2allphase}.

The overall structure of polarized light from model B2 is similar to the simplified model, shown in Fig.~\ref{simplemodel}, with bright arcs where the density of the dust is high, and decreasing intensity closer to the star. The most notable difference between Fig.~\ref{B2allphase} and Fig.~\ref{simplemodel} is the non-zero intensity outside the bright arcs in Fig.~\ref{B2allphase}. This is due to the more extended distribution of dust in the detailed model compared to the geometrically thin dust shell in the simple test case. Clear variations in the polarized light with pulsation phase can also be seen in Fig.~\ref{B2allphase}, both regarding the overall intensity and the radial position of the arcs. 

This variation in polarized light is caused by temporal changes in grain size and density structure during the pulsation cycle, as illustrated in Fig.~\ref{densityPlot}. A new dust layer starts to condense at the inner edge of the existing dust shell around the luminosity minimum (panel 3, $\phi=0.5$) when the overall temperature of the atmosphere is lowest. In panel three of Fig.~\ref{densityPlot}, a small peak in dust density appears with grain sizes around $0.15\,\mu$m. During the consecutive phases the newly formed dust layer moves outwards and the grains grow, reaching a size of $0.45\,\mu$m near the luminosity maximum (see panel 1, $\phi=0.00$, which follows panel 4 because of the cyclic behaviour of the model). The change in the radial position of the density peak is reflected in the location of the arcs in Fig.~\ref{B2allphase}.

The top left panel of Fig.~\ref{B2allphase} (phase 0 at $1.2\,\mu$m) shows another interesting phenomenon, namely a dependence of polarization state on the distance from the star. This is caused by the fact that light seen at different distances has been scattered at different angles and that the polarization state depends on the scattering angle. The angle, and thereby the distance, at which this change in polarization state occurs depends on the size of the dust grains relative to the wavelength of the light. This polarization flip can be seen in Fig.~\ref{B2allphase} for $\lambda=1.2\,\mathrm{\mu m}$ as a change in colour from red to blue (or vice versa) with radial distance. Light scattered at all angles has the same state of polarization at longer wavelengths ($\lambda=1.5\,\mathrm{\mu m}$ and $\lambda=2.1\,\mathrm{\mu
	m}$). In principle, the detection of arcs with a flip in their polarization state could be used as a way to determine grain sizes, although it requires very good spatial resolution. 

\subsection{Simulating non-sphericity}
\label{nonsphere}
Up to this point we only considered spherically symmetric models, assuming the same properties of the dust and gas shell in all directions. However, imaging of circumstellar dust close to AGB stars shows significant deviations from spherical symmetry. Recently, \citet{ohnaetal16a, ohnaetal16b} and \citet{khouetal16} have obtained direct images of polarized light around two nearby AGB stars (W Hya and R Dor) with SPHERE/ZIMPOL, demonstrating the variable, non-spherical nature of the circumstellar material. The images of both stars show clumpy structures, which vary on timescales of several months. In particular, \citet{ohnaetal16a, ohnaetal16b} have interpreted the observed change in dominant grain size between pre-maximum and minimum phases of W Hya as an effect of grain growth during the phases of increasing luminosity, following the formation of new grains around the luminosity minimum. Both, the grain sizes, of about 0.1 and 0.5 micron at minimum and pre-maximum phases, respectively, and the timing of grain growth derived from these observations, fit very well with our model B2, as shown in Fig.~\ref{densityPlot}.     

\begin{figure}
	\centering
	\includegraphics[width=\hsize]{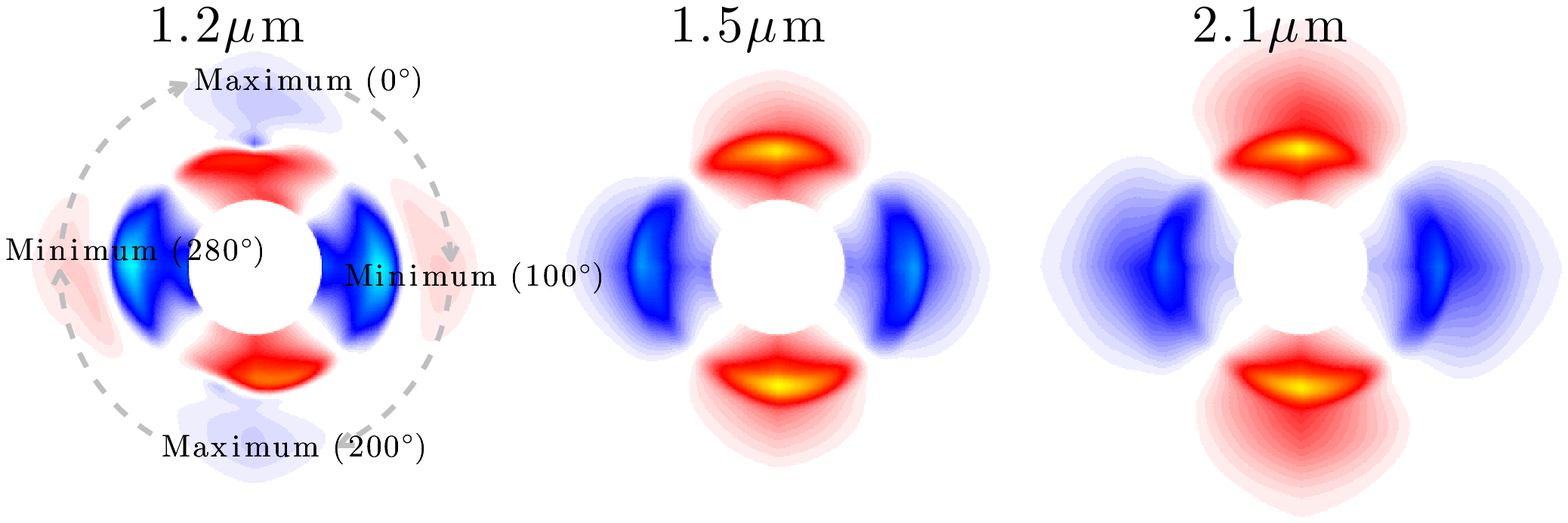}
	\includegraphics[width=\hsize]{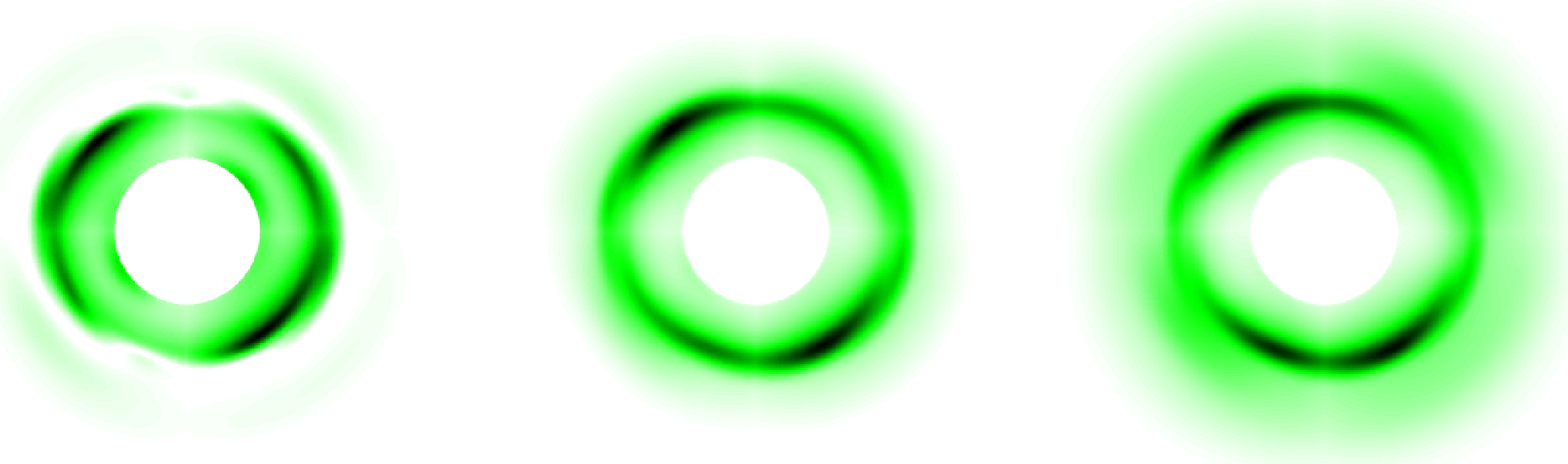}
	\caption{Non-spherically symmetric version of model B2 (composite of different phases) in polarized light. \newline
		The upper panel represents the horizontally and vertically polarized light; the colour coding is the same as in Fig.~\ref{simplemodel}.  The lower panel represents the combined polarized intensity.}
	\label{B2nonSphere}
\end{figure}

To mimic a non-spherical structure, as in the observations, we combine a range of pulsation phases into a single synthetic image, using snapshots from model B2 as a radial structure in a certain direction, rather than a global, spherically symmetric structure. Two maximum phases are placed at opposite sides of the star, but slightly of-axis (at $0^{\circ}$ and $200^{\circ}$). Intermediate phases are spread evenly between the maximum phases in clockwise direction with interpolation between phases to create a smooth transition. See left panel of Fig.~\ref{B2nonSphere} for a sketch of where phases are located. The middle and right panels of this figure show the polarized light for the same underlying structure but at different wavelengths. A difference compared to a spherically symmetric model is that the images now produce a net polarization.

\subsection{Dust size diagnostics} 
\label{Dust size determination}
\begin{figure}
	\centering
	\includegraphics[width=\hsize]{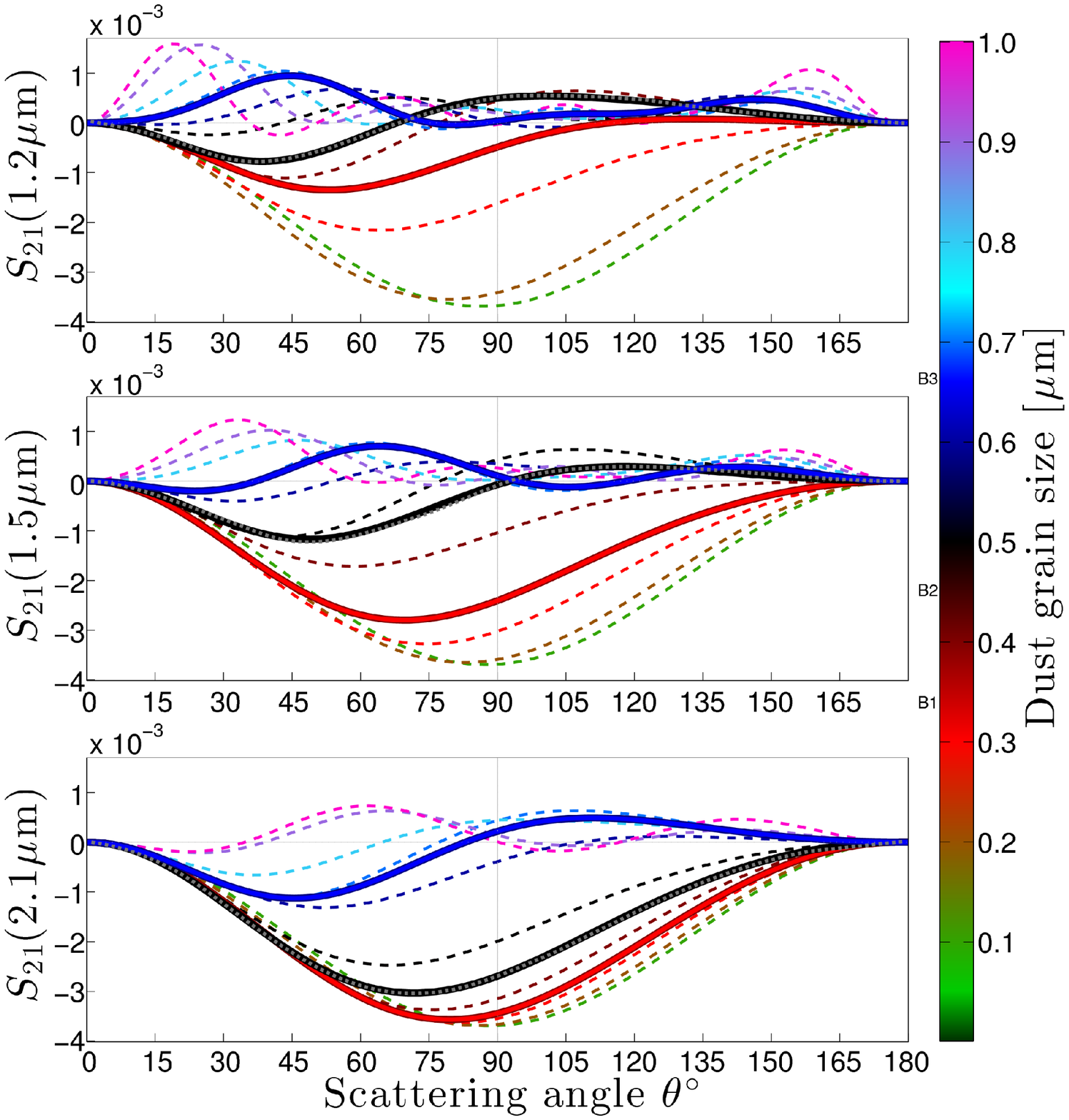}
	\caption{Matrix element $S_{21}$ of Mie-scattering matrices for Mg$_2$SiO$_4$ dust grain with optical data from \citet{Jager2003JQSRT}. The top panel indicates $\lambda = 1.2\,\mathrm{\mu m}$, middle panel indicates $\lambda = 1.5\,\mathrm{\mu m}$, and bottom panel indicates $\lambda = 2.1\,\mathrm{\mu m}$. The dashed lines show $S_{21}$ for a distribution of dust grain sizes from 0.1 to $1.0\,\mathrm{\mu m}$, and the solid lines show $S_{21}$ for sizes corresponding to the average size in models. Grey dotted lines show the same quantity calculated for 0.45 $\mathrm{\mu m}$ sized forsterite grains using the light scattering code MiePlot.}
	\label{S21_all}
\end{figure}

Grain size is a critical parameter, which determines the contribution of pure scattering to the radiative acceleration experienced by the dust, and therefore has important consequences for the wind mechanism. The state of polarization of scattered light depends on both the size parameter (i.e. the relative size of scattering particles to wavelength of light) and the scattering angle. Scattering matrix elements $S_{21}$ (shown in Fig.~\ref{S21_all}) gives the intensity and state of polarization of scattered light, where the negative values correspond to polarization perpendicular to direction of incoming light and positive values correspond to polarization parallel to incoming light, as function of scattering angle. As the size parameter changes, the state of polarization for a given scattering angle can change. For model B2, with dust grains with radius around $0.45\,\mathrm{\mu m}$ (solid black line in Fig.~\ref{S21_all}), $S_{21}$ changes from positive to negative at scattering angles around $90^\circ$ when the wavelength increases.

\begin{figure*}
	\centering
	\includegraphics[width=\hsize]{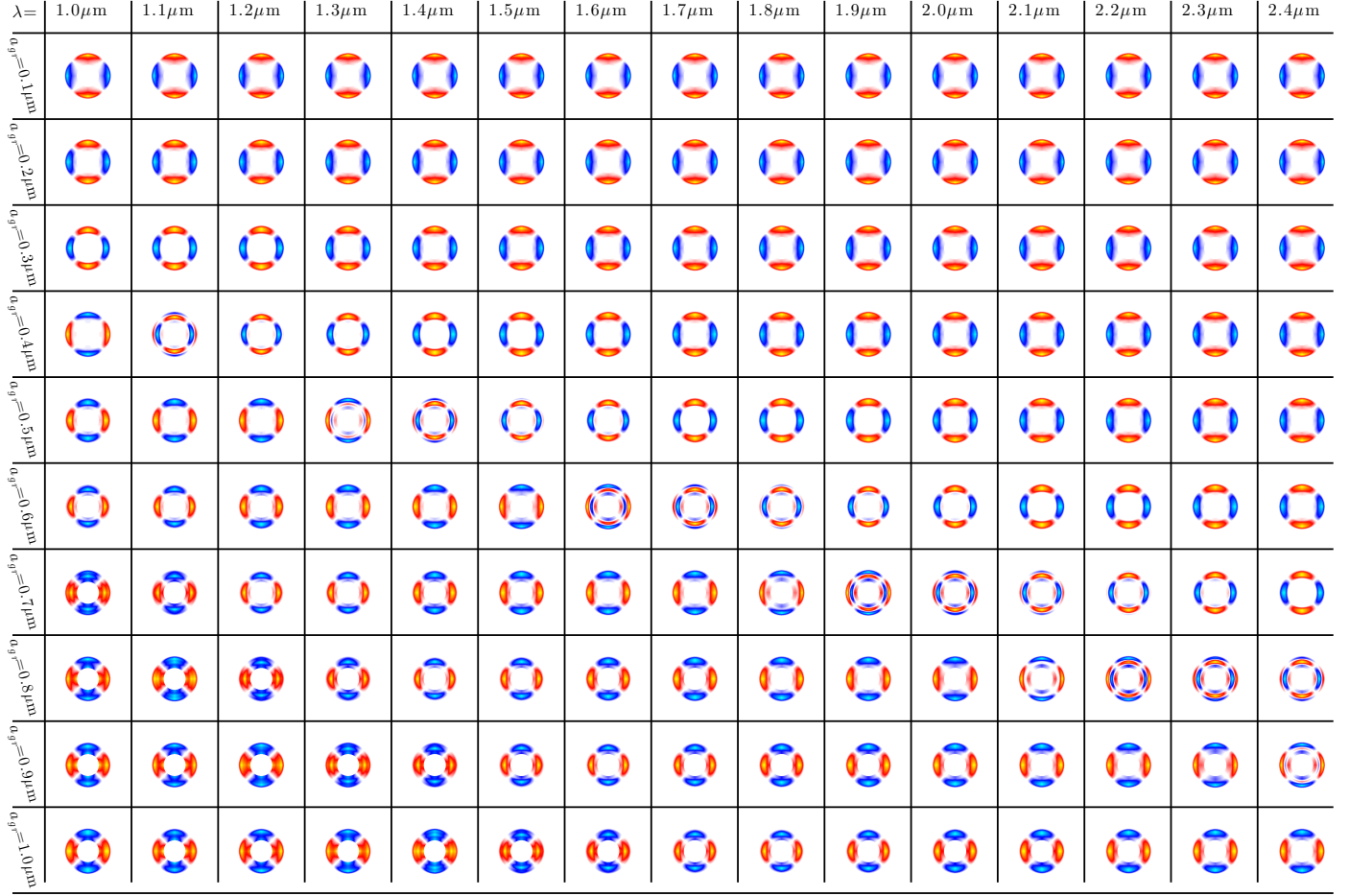}
	\caption{Linearly polarized light from a grid of simplified models. Colour coding is the same as in Fig.~\ref{simplemodel}.}
	\label{allSimple}
\end{figure*}

To see how the state of polarization of scattered light changes with grain size and wavelength, we calculate the polarized light for a set of simplified models with a thin spherically symmetric dust shell situated at $2.5\,R_{\star}$,  consisting of dust particles of uniform size (see Fig.~\ref{allSimple}). Synthetic polarized light was computed for 10 different dust sizes and in wavelengths ranging from 1.0 to $2.5\,\mu$m. In the wavelength region close to where the state of polarization switches from vertical to horizontal, at scattering angles around 90$^\circ$ (i.e. from positive to negative in Fig.~\ref{S21_all}), one can observe multiple bright arcs of polarized light with alternating states of polarization. At shorter wavelengths than this threshold, the polarization is dominated by light polarized parallel to local stellar radiation, while at longer wavelengths, polarization is dominated by light polarized perpendicular to local stellar radiation. In all cases, the switch from one dominant state of polarization to the other happens at wavelengths around three times the radius of dust grains. Thus, if such change in state of polarization could be observed, the size of the scattering particles could easily be estimated.

\begin{figure*} 
	\centering
	\includegraphics[width=\hsize]{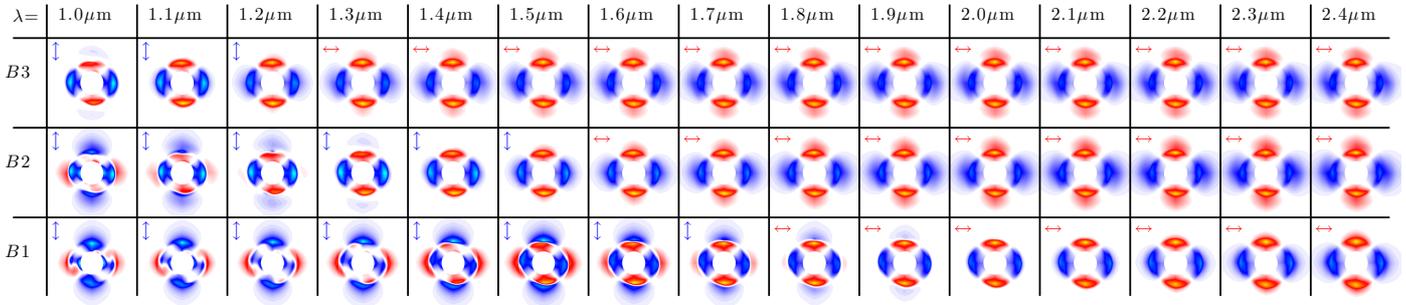}
	\caption{Linearly polarized light from non-spherically symmetric detailed models with varying grain size and at different wavelengths. The dominating state of polarization is indicated by arrows at the top left corner of each model. Colour coding is the same as in Fig.~\ref{simplemodel}. Row B3 shows$\langle a_{gr} \rangle = 0.33\,\mathrm{\mu m}$, row B2 shows $\langle a_{gr} \rangle = 0.45\,\mathrm{\mu m}$ , and row B1 shows $\langle a_{gr} \rangle = 0.66\,\mathrm{\mu m}$.}
	\label{nonSphereB123}
\end{figure*}

To observe changes in state of the polarization, high spatial resolution, sufficient to resolve individual arcs of polarized light, would be needed. Alternately, this could also be observed in unresolved images if the star is asymmetrical in such way that a net polarized signal from the shell could be detected. One such example, based on model B2, was discussed in Sect.~\ref{nonsphere} and shown in Fig.~\ref{B2nonSphere}. In a similar way, we compute images for non-spherical structures constructed from all three models in Tab.~\ref{modelpar}, in wavelengths from 1.0 to $2.5\,\mathrm{\mu m}$; see Fig.~\ref{nonSphereB123}. Along with a spatially resolved image of each model, the net state of polarization of the combined dust shell is indicated with an arrow, assuming no net polarization from the star itself. In all cases net polarization changes orientation at wavelength within $\pm 1\,\mathrm{\mu m}$ as predicted by simplified models; see Fig.~\ref{allSimple}.

\section{Discussion and conclusions}
In this paper we have presented a tool for computing synthetic images of scattered light from detailed dynamical models of gas and dust shells around AGB stars, and we have discussed the diagnostic potential of polarization regarding dust grain sizes. The study presented here was originally inspired by the work of \citet{Norris2012Natur}, who used aperture masking polarimetric interferometry at near-IR wavelengths to determine grain sizes in the close vicinity of W Hya, R Dor, and R Leo. In particular, our intention was to check how assumptions about the radial structure of the circumstellar material, for example regarding density and grain size, affect the resulting scattered light. \citet{Norris2012Natur} used the assumption of a geometrically thin, spherically symmetric shell of dust particles with a single grain size for interpreting their observed data. In contrast, our results are based on extended, time-dependent dynamic atmosphere and wind models, computed with the DARWIN code, which feature a detailed treatment of grain growth \citep{Hofner2008AA, Bladh2013AA, Hoefner2016AA}

From the corresponding synthetic images the importance of accounting for temporal changes with pulsation phase becomes apparent. The intensity of the polarized light, the position of the inner edge of the dust shell (where the polarized intensity is highest), and the size of the dust grains near the inner edge are all changing with the luminosity phase, which significantly affects the resulting polarized signal. In addition to the temporal variations caused by radial pulsations, non-spherical structures in the dust shells, for example due to large-scale convective flows  \citep{FreyHoef2008}, can also have an impact on the polarized light. In the present paper, we simulated this effect by combining a range of pulsation phases of a 1D model into a single 3D structure, before computing synthetic images. We therefore conclude that the interpretation of data using time-independent, spherically symmetric models should be made with caution.

While the complex, variable morphology of the circumstellar dust shell can make the interpretation of resolved data for nearby stars more difficult, it might turn out to be a blessing in disguise for unresolved objects. An asymmetry of the circumstellar envelope creates a net polarization of the total scattered light from the dust shell, which could give diagnostics of grain sizes. The ratio between the size of the scattering particles and the observed wavelength is the dominant parameter determining at what wavelengths net polarization switches from parallel to perpendicular (see Fig.~\ref{S21_all}). By determining at what wavelength the dominant state of polarization changes, average particle size could be constrained. It should, however, be noted that a non-detection of such a change in net polarization cannot be used to put constraints on grain sizes. Apart from grains not being within the corresponding size regime for the given wavelength range, there could be numerous explanations, for example, no significant deviations from spherical symmetry, superposition of grains with a large spread in sizes, polarized light dominated by gas from the inner parts of the shell, etc.

Finally, it should be mentioned that the chemical composition of the dust grains cannot be reliably determined from the scattered light observations at visible and near-IR wavelengths alone, but only in combination with measurements in the mid-IR regime where the solid particles show characteristic lattice modes. Considerations of thermal stability and radiative heating in the close vicinity of the star suggest Fe-free silicates (Mg$_2$SiO$_4$, MgSiO$_3$) and/or Al$_2$O$_3$ as likely candidates for the light-scattering dust grains. Spectro-interferometric observations of AGB stars indicate that such grains can exist around 2 stellar radii, or even below, with Al$_2$O$_3$ forming closer to the star than silicates (e.g. \citet{wittetal07}, \citep{zhaoetal11, zhaoetal12}, \citet{karoetal13}, \citet{sacuetal13}). This is consistent with an interpretation that the smaller grains found by \citet{ohnaetal16b} around W Hya near minimum light may consist of Al$_2$O$_3$ and that they grow silicate mantles at later phases. The scenario of winds driven by grains consisting of Al$_2$O$_3$ cores and Mg$_2$SiO$_4$ mantles was recently studied with DARWIN models by \citet{Hoefner2016AA}, resulting in both, realistic wind properties and variations of visual and near-IR colours in good agreement with observations. In this context, the Al$_2$O$_3$ particles prior to silicate condensation experience a radiative pressure that is insufficient for driving an outflow, and they form a gravitationally bound dust shell close to the star in accordance with the results of \citet{ireletal05} and \citet{khouetal15, khouetal16}. 

\begin{acknowledgements}
	This work has been supported by the Swedish Research Council (\textit{Vetenskapsr\aa det}) and by the ERC Consolidator Grant funding scheme (project {\em project STARKEY}, G.A. No.~615604). The computations of spectra and photometry were performed on resources provided by the Swedish National Infrastructure for Computing (SNIC) at UPPMAX.
\end{acknowledgements}

\bibliography{mybib12.bib} 
\bibliographystyle{aa}

\end{document}